\documentstyle[preprint,aps]{revtex}
\begin{document}
\draft
\title{On the problem of neural network decomposition into some subnets}
\author{Leonid B. Litinskii}
\address{Institute of High Pressure Physics of Russian Academy of Sciences\\
Russia, 142092 Troitsk Moscow region, e-mail: u10740@dialup.podolsk.ru}
\maketitle
\begin{abstract}
An artificial neural network is usually treated as a whole system, being
characterized by its ground state (the global minimum of the energy
functional), the set of fixed points, their basins of attraction, etc. However,
it is quite natural to suppose that a large network may consist of a set of
almost autonome subnets. Each subnet works independently (or almost
independently) and analyzes the same pattern from other points of view. 
It seems that it is a proper model for the natural neural networks. We discuss
the problem of decomposition of a neural network into a set of weakly coupled
subnets. The used technique is similar to the method for {\it the extremal
grouping of parameters}, proposed by E.M.Braverman (1970).
\end{abstract}
\section{Hopfield's model of a neural network}
A neural network of size $n$ is a set of $n$ connected spin variables 
({\it spins}) $\sigma_i$; each $\sigma_i$ can be either $1$ or $-1$:
$$\sigma_i = {\{\pm 1\}},\quad i=1,2,...n.\eqno(1)$$
The interaction between spins is described by a connection matrix. Let 
$J_{ii'}$ be the connection strength between the spins $\sigma_i$ and
$\sigma_{i'}$\footnote{For the sake of simplicity we suppose that there is no
self-interaction in the system: $J_{ii}=0\ \forall i$.}, and 
let $\sigma_i(t)$ be the value of $i$th spin at time $t$, then
$$h_i(t) = \sum^n_{i'=1}J_{ii'}\cdot \sigma_{i'}(t)\eqno(2)$$
represents the local field that the spin $\sigma_i$  experiences at  time $t$. 
Under the action of this field the new value of the spin $\sigma_i$ at  the 
next moment $t+1$ is:
$$\sigma_i(t+1) = \left\{\begin{array}{rl}\sigma_i(t),&\mbox{ if }
h_i(t)\cdot\sigma_i(t)\ge 0\\
-\sigma_i(t),&\mbox{ if }
h_i(t)\cdot\sigma_i(t)< 0\end{array}\right.\eqno(3)$$
The vectors which coordinates are $\{\pm 1\}$ only is called the 
{\it configuration
vectors}. We denote the configuration vectors by small Greek letters. 

It is convenient to describe the state of the network at time $t$ by 
$n$-dimensional configuration vector 
$$\vec\sigma(t)=(\sigma_1(t),\sigma_2(t),\ldots,\sigma_n(t)).$$
If we introduce the  connection  matrix  ${\bf J} = \left(J_{ii'}\right)_1^n$
and define the quadratic form
$$E(t) = -\sum^n_{i,i'=1}J_{ii'}\cdot\sigma_i(t)\cdot\sigma_{i'}(t) = 
-({\bf J}\vec\sigma(t),\vec\sigma(t)),\eqno(4)$$
then it is easy to show that for any \underline{symmetrical} connection matrix
$\bf J$ the  
overturn of a spin $\sigma_i(t)$, which value does not coincide with the sign
of $h_i(t)$, leads to the decrease of $E(t)$:
$$E(t+1) = E(t) + 4\cdot\sigma_i(t)\cdot h_i(t).\eqno(5)$$
$E(t)$ can be interpreted as the energy of the state $\vec\sigma(t)$. 
As the number 
of network states is finite and the $i$th spin does not turn over if
$h_i(t) = 0$, it is obvious that the final state of the network would be a 
state which corresponds to a minimum (may be local) of the energy $E(t)$. 
In such a state every spin $\sigma_i$  will be align with its local field 
$h_i$  
and there will be no further evolution of the network. These states are 
called the {\it fixed points} of the network. Consequently, if the 
configuration vector $\vec\sigma^*=(\sigma^*_1,\sigma^*_2,\ldots,\sigma^*_n)$
is a fixed points, then  
$$\sigma^*_i = sgn\left(\sum_{i'=1}^nJ_{ii'}\cdot\sigma^*_{i'}\right),\quad 
i=1,2,\ldots,n.\eqno(6)$$
In what follows the configuration vectors which are fixed points will be 
marked by superscripts "*".

Let's define a neural network which is called Hopfield's network. Let 
$p$ be a number of preassigned configuration vectors $\vec\xi^{(l)}$, which
are called the {\it memorized patterns}:
$$\vec\xi^{(l)}=(\xi^{(l)}_1,\xi^{(l)}_2,\ldots,\xi^{(l)}_n),\quad
l=1,2,\ldots,p.\eqno(7)$$ 
(The superscripts numerate  the  vectors  from  $\rm R^n$ and  the  subscripts 
numerate their coordinates. Usually it is assumed that $p<n$ or even $p<<n$.) 
J.Hopfield\cite{Hopf} proposed to use the connection matrix of the form:
$$J_{ii'}=\left\{\begin{array}{ll}\sum_{l=1}^p\xi_i^{(l)}\xi_{i'}^{(l)},&
i\ne i'\\0,&i=i',\quad i,i'=1,2,\ldots,n.\end{array}\right.\eqno(8)$$        
The matrix ${\bf J}$ (8) is a symmetric matrix with zero diagonal elements.
Then, the fixed points are the minima of the energy $E$ given by Eq.(4). If
we define $(p\times n)$-matrix $\bf\Xi$ with $p$ memorized patterns (7) as the
rows, 
$${\bf\Xi}=\left(\begin{array}{c}\vec\xi^{(1)}\\\vec\xi^{(2)}\\\vdots\\
\vec\xi^{(p)}\end{array}\right)=\left(\begin{array}{cccc}
\xi^{(1)}_1&\xi^{(1)}_2&\ldots&\xi^{(1)}_n\\
\xi^{(2)}_1&\xi^{(2)}_2&\ldots&\xi^{(2)}_n\\
\vdots&\vdots&\ldots&\vdots\\
\xi^{(p)}_1&\xi^{(p)}_2&\ldots&\xi^{(p)}_n\end{array}\right)\eqno(9)$$
then the expression for the connection matrix takes the form                  
$${\bf J}= {\bf\Xi}^T\cdot{\bf\Xi} - p\cdot{\bf I},\eqno(10)$$
where $(n\times p)$-matrix ${\bf\Xi}^T$ is the transpose of matrix $\bf\Xi$ and
$\bf I$ is the unit matrix in the space $\rm R^n$. Therefore the searching of
the fixed points of Hopfield's network reduces to the maximization of the
functional 
$$({\bf\Xi}^T\cdot{\bf\Xi}
\vec\sigma,\vec\sigma)=\parallel{\bf\Xi}\vec\sigma\parallel^2.$$
But this problem can be
reformulated, if $n$ $p$-dimensional vectors $\vec\xi_i$,  
which are \underline{the columns} of matrix $\bf\Xi$ are introduced:
$$\vec\xi_i= \left(\begin{array}{c}\xi_i^{(1)}\\\xi_i^{(2)}\\\vdots\\
\xi_i^{(p)}\end{array}\right)\in {\rm R}^p,\quad i=1,2,\ldots,n.\eqno(11)$$
In contrast to $n$-dimensional vectors $\vec\xi^{(l)}$ defined by Eq.(7),
here the subscripts numerate the vectors $\vec\xi_i$ from $\rm R^p$ and the
superscripts numerate their coordinates.

It is easy to see, that the problem of maximization of the functional 
$\parallel{\bf\Xi}\vec\sigma\parallel^2$ takes the form:
$$\parallel\sum_{i=1}^n\sigma_i\cdot\vec\xi_i\parallel\to\mbox{ max},\ 
\mbox{ where } \sigma_i=\{\pm 1\}\ \forall i.\eqno(12)$$
In other words, we have to find out such a weighted sum of the $p$-dimensional 
vectors $\vec\xi_i$  with the weights are equal $\{\pm 1\}$, which length would
be maximal. In what follows the expression (12) would be the start point of our
consideration. 

\section{Factor analysis and extremal grouping of parameters}

The problem (12) is a special case of the problem which is well-known 
for the centroid method of the factor analysis\cite{Harm}.
The basic idea of the factor analysis is to replace the great  number 
of the parameters, which describe the objects under investigation, by  a 
considerably lesser set of specially constructed characteristics provided 
that such replacement would not lead to the loss of the essential 
information about these objects.

The formalization of this idea can be done in the following way.
Let us have $p$ objects which are represented by the vectors 
$\vec x^{(l)} = (x^{(l)}_1,x^{(l)}_2,\ldots,x^{(l)}_n),\ l=1,2,\ldots,p$ 
in the space $\rm R^n$. Let's consider the 
$(p\times n)$-matrix $\bf X$, which rows are the object-vectors $\vec x^{(l)}$.
(This matrix is an analog of the matrix $\bf\Xi$ (9), but now the matrix
elements  
can be an arbitrary real numbers, and not $\pm 1$ only.) On the other hand
the matrix $\bf X$ can be described as the matrix which columns are the 
parameter-vectors $\vec x_i$: 
$$\vec x_i= \left(\begin{array}{c} x^{(1)}_i\\x^{(2)}_i\\\vdots\\x^{(p)}_i
\end{array}\right),\quad i=1,2,..n.$$ 
(We recall that the vectors from the space $\rm R^n$ are numerated by 
superscripts: $l=1,\ldots,p$, and the vectors from the space $\rm R^p$ 
by subscripts: $i=1,\ldots,n$.)

If a relatively small number $t$ ($t<<n$) of such $p$-dimensional 
vectors $\vec f_1,\vec f_2,\ldots,\vec f_t$  can be found, that 
the papameter-vectors $\vec x_i$ can be represented in the form 
$$\vec x_i=\sum_{s=1}^t a_{is}\cdot\vec f_s + \vec a_i,\quad 
i=1,2,\ldots,n,$$
where the remainders $\vec a_i$  are small in some sense and can be omitted, 
then the objects can be described by the characteristics $\vec f_s$ instead 
of the initially used parameters $\vec x_i$. Indeed, due to the smallness of
the remainders $\vec a_i$, characteristics $\vec f_s$ adequately describe the
investigated phenomenon. But it is much more convenient to work if the number
of the parameters is considerable reduced. The characteristics $\vec f_s$ are
called {\it the essential factors}.

The various models of the factor analysis differ in the forms in which 
the factors $\vec f_s$  are sought and the sense in which the smallness of 
$\vec a_i$ is understood. In the centroid method {\it the first} factor
$\vec f_1$ is sought as a linear combination 
$\sum_{i=1}^n\sigma_i\cdot\vec x_i$ of the parameters $\vec x_i$ with the
weights $\sigma_i=\{\pm 1\}$, that have a maximal length
$$\vec f_1\propto\sum_{i=1}^n\sigma^*_i\cdot\vec x_i,\mbox{ where }
\parallel\sum_{i=1}^n\sigma^*_i\cdot\vec x_i\parallel
=\max_{\sigma_i=\{\pm 1\}}\parallel\sum_{i=1}^n\sigma_i\cdot\vec x_i\parallel
.\eqno(13)$$
The comparison of Eq.(12) and Eq.(13) shows that the problem of the 
network fixed points searching is equivalent to the construction of the 
first centroid factor for the set of the $p$-dimensional vectors $\vec\xi_i$
(11). 

In the centroid method after the construction of the first factor $\vec f_1$,
the vectors $b_{i1}\cdot\vec f_1$, where $b_{i1},\ i=1,2,.,n$  are some
coefficients, are 
subtracted from each parameter-vector $\vec x_i$. In such a way we obtain a
new set of vectors $\vec x_i^\prime = \vec x_i - b_{i1}\cdot\vec f_1$  for
which their own factor is constructed  
by analogy. This factor would be the second factor for the 
initial 
parameters $\vec x_i$ . This process will be repeated till the vectors which
are obtained after the next step would be small enough. For details see 
\cite{Harm,Brav1,Brav2}.

An important generalization of the factor analysis was the idea of 
the {\it extremal grouping of the parameters} suggested by E.M.Braverman in 
1970\cite{Brav1}. Braverman introduced a model of the factor analysis where
an essentially nonuniform distribution of the vectors $\vec x_i$ in the space 
$\rm R^p$ was taken into account. 

Indeed, if the number $n$ of the parameter-vectors is very large, it is 
possible that they can be divided into some compact groups such that the 
vectors joined into one group are "strongly correlated" with each other 
and are "weakly correlated" with  the  parameters  included  into  other 
groups. Then it is reasonable to construct the factors not for the full 
set of the parameter-vectors, but for every compact group separately. If
these groups are compact enough, we  can  restrict  ourselves  with  the 
first factor of each group only. To divide the parameter-vectors into 
these compact groups, Braverman suggested an approach connected with the 
maximization of a certain functional depending both on the  
grouping of the parameters and on the choice of the factors.

Let's write down Braverman's functional. Let $p$-dimensional vectors 
$\vec x_1,\vec x_2,\ldots,\vec x_n$
be divided into current disjoint groups $A_1, A_2,\ldots, A_t$:
$$A_1\bigcup A_2\bigcup\ldots\bigcup A_t=\{1,2,\ldots,n\}.$$ 
For every group $A_s$ the first centroid factor can be constructed as the
solution of the problem:
$$\parallel\sum_{i\in A_s}\sigma^*_i\cdot\vec x_i\parallel
=\max_{\sigma_i=\{\pm 1\}}\parallel\sum_{i\in A_s}\sigma_i\cdot\vec x_i
\parallel.\eqno(14)$$
Then, the partition into $t$ the most compact groups is obtained as a result 
of the maximization of the functional:
$$M(A_1,A_2,\ldots,A_t)= 
\parallel\sum_{i\in A_1}\sigma^*_i\cdot\vec x_i\parallel+
\parallel\sum_{i\in A_2}\sigma^*_i\cdot\vec x_i\parallel+\ldots +
\parallel\sum_{i\in A_t}\sigma^*_i\cdot\vec x_i\parallel\to\max\eqno(15)$$
where $\sigma^*_i$ are the solutions of the problem (14) for every group
$A_s,\quad s =1,2,..,t$.
We want to notice, that, though the problem of maximization of the functional
(15) is very hard, 
the  method  for  the  extremal  grouping  of 
parameters was successfully used for various problems in engineering, 
economics, sociology, psychology and other fields \cite{Brav1,Brav2}.

\section{Neural networks decomposition into some subnets}
Let in Eqs.(14),(15) the vectors $\vec x_i$  be replaced by the vectors 
$\vec\xi_i$ from Eq.(11), i.e. only the vectors with the coordinates 
$\{\pm 1\}$ are under consideration. Then, in the framework of the neural
network paradigm,  the  
problem (14),(15) can be interpreted as the problem of the  grouping  of 
the network neurons into some connected groups.

Indeed,  natural  networks  have  evident   differential   structure: 
different neuron groups have different functions, they respond  for  the 
regulation/analysis of different aspects of a complicate pattern 
which is worked over by the network. To some extent  every  such  neuron 
group can be treated as an autonomous neural network of the smaller size 
which is dealing with some specific features of the pattern.

Let a network be consisted of some groups of neurons (subnets)
$A_1, A_2,\ldots,A_t$. There is one universal mechanism for the functioning of 
all network neurons: a spin $\sigma_i$  turns over if its sign does not
coincide with the sign of the field $h_i$ acting on this spin. However, it is 
reasonable to assume that the incoming excitations from the  neurons 
belonging to the same group as the neuron $\sigma_i$ affect this neuron
stronger  
then the excitations from the neurons of other groups (those, which 
analyze the same pattern from other points of view). This hierarchy of 
excitations can be modelled in different ways. As an initial model it 
can be assumed that:
$$h_i(t)=\sum^n_{i'\in A_s}J_{ii'}\cdot \sigma_{i'}(t)\eqno(16)$$
where the summation is taken over all neurons belonging to the same 
group $A_s$ as the $i$th neuron.

The subnet consisting of the neurons from the group $A_s$ is evolving to 
one of its fixed points. This leads us to the problem (14). And the network 
as a whole is acting so, that the composite functional
$$M(\{\sigma_i\}_1^n)= 
\parallel\sum_{i\in A_1}\sigma_i\cdot\vec\xi_i\parallel+
\parallel\sum_{i\in A_2}\sigma_i\cdot\vec\xi_i\parallel+\ldots +
\parallel\sum_{i\in A_t}\sigma_i\cdot\vec\xi_i\parallel\to\max\eqno(17)$$
would be maximized.

We have discussed the situation when the neurons are already 
decomposed into groups $A_1,A_2,\ldots,A_t$. If the structure of the groups is
unknown,  
but their number $t$ is fixed, it is necessary to maximize the functional 
(17) with respect to the structure of the groups $A_s$ as well as with 
respect to all the weights $\sigma_i$ inside every group. In this case 
Eqs.(14),(15), where the vectors $\vec\xi_i$ have to be substituted instead of 
the vectors $\vec x_i$, describe the optimal decomposition of the network into
$t$ autonomous subnets. 

Here some remarks must be done.
Firstly, it is easy to see, that when the number of the groups $t$
increases, the functional $M(A_1,..A_t)$ (15) is nondecreasing (it follows 
from the triangle inequality). This functional attains it's global 
maximum when the number of the groups $t$ is equal $n$. However, it is a
trivial decomposition. Simple geometric arguments show that when a group of 
\underline{strongly} correlated vectors $\vec x_i$ is divided into two
subgroups, the  
functional (15) increases negligibly. So, the problem is not to get the 
global maximum of the functional (15), but to obtain such a number $t^*$ of 
the groups beginning with which the further 
increase of the number of the groups would not lead to the substantial 
increase of this functional. About these $t^*$ groups we can speak as about 
{\it the proper} number of the subnets which constitute the initial
$n$-network.  

Furthermore, it is reasonable to try to interpret the  specific 
characteristics of each obtained subnet in meaning terms. In other 
words, we can try to understand what kind of pattern's characteristics 
are analyzed by each particular subnet, i.e. we must determine what kind 
of neurons are joined in the group. On this step the monograph \cite{Brav2}, 
which reflects the accumulated experience in this field, can be useful.

Secondly, the above mentioned program can be fulfilled only if we are 
able to solve two problems: A) to find out the compact  groups  of  the 
vectors $\vec\xi_i$; B) to determine the optimal configuration
$\{\sigma_i^*\},\  
i\in A_s$, for each group. What concerns the problem B, actually all the
attempts to create an effective algorithm for the maximization of the
functional $({\bf J}\vec\sigma,\vec\sigma)$ are devoted to this problem.

The problem A is much less studied and seems to be more complicated.
Usually, it is solved by step  
by step transferring of $p$-dimensional vectors from one group to another, 
and the comparison  of  the  values  of  the  functional  (15)  for  the 
consequently obtained grouping. When $n$ is rather large, in  such  a  way 
only the determination  of  the  local  maximum  of  the  functional  $M$ is 
guaranteed. 
We know not so much papers \cite{Bukh,Lit}, devoted exactly to the problem of
finding of the global maximum of a functional  of type (15).
In these papers the general case of vectors $\vec x_i$
with {\it real} coordinates is studied. As for neural networks, the 
vectors $\vec\xi_i$ are specific: their coordinates are $\{\pm 1\}$. It can be 
hope that the specific character of the vectors $\vec\xi_i$  would make it
possible to present effective method for the searching of the compact
groups.  

And the last remark. Although  the  proposed  approach  was  formulated  for 
Hopfield's model, it can be generalized for  the  case  of  an  arbitrary 
symmetric connection matrix: it is sufficient to replace in Eqs. (14), (15) and
(17) the term
$$\parallel\sum_{i\in A_s}\sigma^*_i\cdot\vec x_i\parallel$$
by
$$\left(\sum_{i,i'\in A_s}J_{ii'}\cdot\sigma_i\cdot\sigma_{i'}\right)^{1/2},$$
and all reasoning are valid. 

This project was supported partially by Russian Basic Research Foundation.

\end{document}